\documentclass[sigconf,10pt]{acmart}

\usepackage[english]{babel}
\usepackage[utf8]{inputenc}
\usepackage{blindtext}
\usepackage{booktabs,caption}
\usepackage{threeparttable}
\usepackage{subfigure}
\usepackage{tumcolor}
\usepackage{amssymb}
\usepackage{pifont}
\newcommand{\cmark}{\textcolor{TUMDarkGreen}{\ding{51}}}%
\newcommand{\xmark}{\textcolor{TUMDarkRed}{\ding{55}}}%

\setcopyright{acmcopyright}

\acmDOI{}

\acmISBN{}

\acmConference{ANCS'19}
\acmYear{2019}
\copyrightyear{2019}

\acmPrice{}

\begin{document}
\title[The Case for Writing Network Drivers in High-Level Programming Languages]{The Case for Writing Network Drivers in\\High-Level Programming Languages}


\author{Paul Emmerich$^{1}$, Simon Ellmann$^{1}$, Fabian Bonk$^{1,}$\footnotemark[1], Alex Egger$^{1,}$\footnotemark[1], Esaú García Sánchez-Torija$^{1,2,}$\footnotemark[1], Thomas Günzel$^{1,}$\footnotemark[1], Sebastian Di Luzio$^{1,}$\footnotemark[1], \mbox{Alexandru Obada$^{1,}$\footnotemark[1]}, Maximilian Stadlmeier$^{1,3,}$\footnotemark[1], Sebastian Voit$^{1,}$\footnotemark[1], Georg Carle$^{1}$\\
\vspace{-0.5em}\normalsize$^1$ Technical University of Munich\\
$^2$ Universitat Politècnica de Catalunya\\
$^3$ Ludwig-Maximilians-Universität München
\normalsize\{emmericp~\textbar~ellmann~\textbar~bonkf~\textbar~eggera~\textbar~garciasa~\textbar~guenzel~\textbar~luzio~\textbar~aobada~\textbar~mstadlme~\textbar~voit~\textbar~carle\}@net.in.tum.de}

\authornote{Equal contribution, ordered alphabetically}

\renewcommand{\shortauthors}{Paul Emmerich et al.}

\begin{abstract}
Drivers are written in C or restricted subsets of C++ on all production-grade server, desktop, and mobile operating systems.
They account for 66\% of the code in Linux, but 39 out of 40 security bugs related to memory safety found in Linux in 2017 are located in drivers.
These bugs could have been prevented by using high-level languages for drivers.
We present user space drivers for the Intel ixgbe 10\,Gbit/s network cards implemented in Rust, Go, C\#, Java, OCaml, Haskell, Swift, JavaScript, and Python written from scratch in idiomatic style for the respective languages.
We quantify costs and benefits of using these languages:
High-level languages are safer (fewer bugs, more safety checks), but run-time safety checks reduce throughput and garbage collection leads to latency spikes.
Out-of-order CPUs mitigate the cost of safety checks: Our Rust driver executes 63\% more instructions per packet but is only 4\% slower than a reference C implementation.
Go's garbage collector keeps latencies below 100\,$\mu$s even under heavy load.
Other languages fare worse, but their unique properties make for an interesting case study.

All implementations are available as free and open source at \href{https://github.com/ixy-languages/ixy-languages}{https://github.com/ixy-languages/ixy-languages}.
\end{abstract}
\maketitle

\section{Introduction}
C has been the go-to language for writing kernels since its inception. 
Device drivers are also mainly written in C, or restricted subsets of C++ providing barely no additional safety features, simply because they are tightly coupled with the kernel in all mainstream operating systems.
Network device drivers managed to escape the grasp of the kernel with user space drivers such as DPDK~\cite{dpdk} in the last years.
Yet, all drivers in DPDK are written in C as large parts of them are derived from kernel implementations.
DPDK consists of more than drivers: it is a whole framework for building fast packet processing apps featuring code from utility functions to complex data structures --- and everything is written in C.
This is not an unreasonable choice: C offers all features required for low-level systems programming and allows fine-grained control over the hardware to achieve high performance.

But with great power comes great responsibility: writing safe C code requires experience and skill.
It is easy to make subtle mistakes that introduce bugs or even security vulnerabilities.
Some of these bugs can be prevented by using a language that enforces certain safety properties of the program. 
Our research questions are: Which languages are suitable for driver development? What are the costs of safety features?
A secondary goal is to simplify driver prototyping and development by providing the primitives required for user space drivers in multiple high-level languages.

\begin{table}[t]
 \setlength{\tabcolsep}{1.2mm}
	\centering
	\footnotesize
	\begin{tabular}{lrrrr}
		\textbf{Language} & \textbf{Main paradigm$^*$} & \textbf{Memory mgmt.} & \textbf{Compilation} \\
		\toprule
		Rust & Imperative & Ownership/RAII$^\dagger$ & Compiled$^\ddagger$ \\
		Go & Imperative & Garbage collection & Compiled \\
		C\# & Object-oriented & Garbage collection & JIT \\
		Java & Object-oriented & Garbage collection & JIT \\
		OCaml & Functional & Garbage collection & Compiled \\
		Haskell & Functional & Garbage collection & Compiled$^\ddagger$ \\
		Swift & Protocol-oriented~\cite{protocol-oriented} & Reference counting & Compiled$^\ddagger$ \\
		JavaScript & Imperative & Garbage collection & JIT \\
		Python & Imperative & Garbage collection & Interpreted \\
		\bottomrule
	\end{tabular}
        \begin{tablenotes}
        		\item \footnotesize $^*$ All selected languages are multi-paradigm
		\item \footnotesize $^\dagger$ Resource Acquisition Is Initialization
		\item \footnotesize $^\ddagger$ Using LLVM
        \end{tablenotes}
	\caption{Languages used by our implementations}
	\label{tbl:languages}
	\vspace{-4em}
\end{table}

We implement a user space driver tuned for performance for the Intel ixgbe family of network controllers (82599ES, X540, X550, and X552) in 9 different high-level languages featuring all major programming paradigms, memory management modes, and compilation techniques.
All implementations are written from scratch in idiomatic style for the language by experienced programmers in the respective language and follow the same basic architecture, allowing for a performance comparison between the high-level languages and a reference C implementation.
For most languages our driver is the first PCIe driver implementation tuned for performance enabling us to quantify the costs of different language safety features in a wide range of high-level languages.
Table~\ref{tbl:languages} lists the core properties of the selected languages.


\section{Related work}
\label{sec:rw}

Operating systems and device drivers predate the C programming language (1972~\cite{ritchie1993development}). 
Operating systems before C were written in languages on an even lower level: assembly or ancient versions of ALGOL and Fortran.
Even Unix started out in assembly language in 1969 before it was re-written in C in 1973~\cite{unix}.
C is a high-level language compared to the previous systems: it allowed Unix to become the first portable operating system running on different architectures in 1977~\cite{unixat25}.
The first operating systems in a language resembling a modern high-level language were the Lisp machines in the 70s (fueled by an AI hype).
They featured operating systems written in Lisp that were fast due to hardware acceleration for high-level language constructs such as garbage collection.
Both the specialized CPUs and operating systems died in the AI winter in the 80s~\cite{aiwinter}.
Operating systems development has been mostly stuck with C since then.

Contemporary related work can be split into three categories:
(1) operating systems and unikernels in high-level languages, (2) packet processing frameworks and user space drivers, and (3) language wrappers making the former available to applications in high-level languages.
Operating systems, unikernels, and language wrappers are discussed with our implementations in the respective languages in Section~\ref{sec:impl}.

Network functions with high performance requirements moved from the kernel into dedicated user space applications in the last decade.
This move happened in two steps: first, frameworks like PF\_RING DNA (2010)~\cite{pfringdna}, PSIO (2010)~\cite{packetshader,psio}, netmap (2011)~\cite{netmap}, and PFQ (2012)~\cite{pfq} provided a fast-path to the network driver from user space applications.
They speed up packet IO by providing a kernel module that maps DMA buffers into a user space application.
These frameworks are not user space drivers: all of them rely on a driver running in the kernel. 
The next step were true user space drivers: DPDK (open sourced in 2013) and Snabb (2012) move the entire driver logic into a user space process by mapping PCIe resources and DMA buffers into a user space library, running the driver in the same process as the application.

An example for this trend is the Click modular router~\cite{click} that started out as a kernel extension in 1999 and was later sped up with a netmap interface in 2012 as a demonstration of netmap itself~\cite{netmap}.
Finally, a DPDK backend was added in 2015 to increase performance even further~\cite{fastclick}.
Similar migration paths can also be found in other open source projects: Open vSwitch comes with a kernel module and had plans to add both netmap and DPDK~\cite{ovs}, the DPDK backend was merged in 2014, the netmap version never left the proof of concept stage~\cite{ovsdpdk}.
pfSense~\cite{pfsense} experimented with both netmap and DPDK in 2015 and finally chose DPDK~\cite{pfsensedpdk}

Applications are moving to user space drivers in the form of DPDK and are therefore free of restrictions imposed by the kernel environment.
Yet, all drivers in DPDK are still written in C.
Snabb~\cite{snabb} (less popular than DPDK and only 4 drivers vs. DPDK's 27 drivers) is the only other performance-optimized user space driver not written in C: It comes with drivers written in Lua running in LuaJIT~\cite{luajit}.
However, it makes extensive use of the LuaJIT foreign function interface~\cite{luajitffi} that erodes memory safety checks that are usually present in Lua.
We are not including Snabb in our performance comparison because its architecture requires ring buffers to connect drivers and ``apps'', this makes it significantly slower than our drivers for the evaluated use case.

\emph{Unrelated work:} Orthogonal to our proposal is the work on XDP~\cite{xdp} as it does not replace drivers but adds a faster interface on top of them.
Moreover, eBPF code for XDP is usually written in a subset of C.
P4~\cite{P4} also deserves a mention here, it is a high-level language for packet processing (but not for the driver itself). It primarily targets hardware, software implementations run on top of existing C drivers.

\section{Motivation}
\label{sec:motivation}
Proponents of operating systems written in high-level languages such as Biscuit (Go)~\cite{biscuit}, Singularity (Sing\#, related to C\#)~\cite{singularity}, JavaOS~\cite{javaos}, House (Haskell)~\cite{house}, and Redox (Rust)~\cite{redox} tout their safety benefits over traditional implementations.
Of these only Redox is under active development with the goal of becoming a production system, the others are research projects and/or abandoned.
These systems are safer, but it is unlikely that the predominant desktop and server operating systems will be replaced any time soon.

We argue that it is not necessary to replace the entire operating system at once.
It is feasible to start writing user space drivers in safer languages today, gradually moving parts of the system to better languages.
Drivers are also the largest attack surface (by lines of code) in modern operating systems and they keep growing in complexity as more and more features are added.
There are real-world security issues in drivers that could have been prevented if they were written in a high-level language.
Moving to user space drivers also means moving to a microkernel architecture with an IPC mechanism between drivers and applications.
User space drivers are more isolated from the rest of the system than kernel drivers and can even run without root privileges if IOMMU hardware is available~\cite{iommu,ixy}.


\subsection{Growing Complexity of Drivers}
\label{sec:driversloc}
66\% of the code in Linux 4.19 (current LTS) is driver code (11.2\,M lines out of 17\,M in total), 21\% (2.35\,M lines) of the driver code is in network drivers.
10 years ago in 2009 Linux 2.6.29 had only 53\% of the code in drivers (3.7\,M out of 6.9\,M lines).
Going back 10 more years to Linux 2.2.0 in 1999, we count 54\% in drivers (646\,k out of 1.2\,M) with only 13\% in network drivers.
One reason for the growing total driver size is that there are more drivers to support more hardware.

\begin{figure}[t!]
     \includegraphics[width=0.45\textwidth]{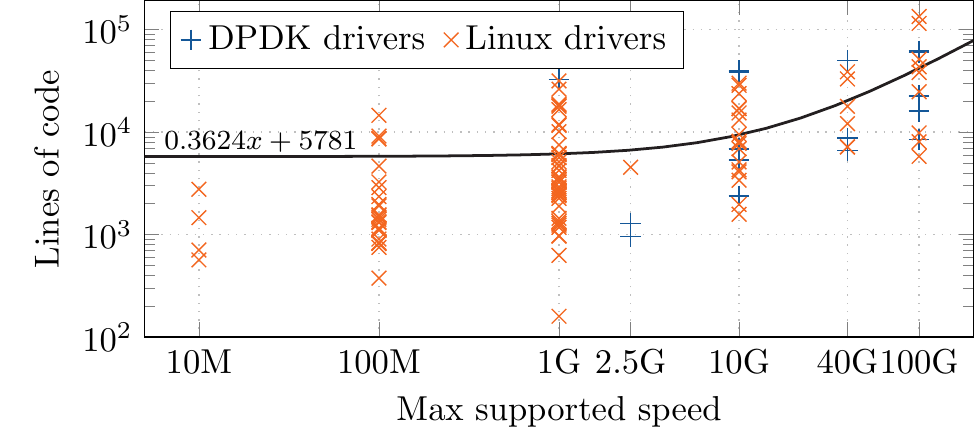}
    \caption{NIC technology node vs. driver size}
    \label{fig:drivers-loc}
\end{figure}

Individual drivers are also increasing in complexity as hardware complexity is on the rise.
Many new hardware features need support from drivers, increasing complexity and attack surface even when the number of drivers running on a given system does not change.
Figure~\ref{fig:drivers-loc} shows a linear correlation ($R^2 = 0.3613$) between NIC technology node and driver complexity as a log-log scatter plot.
The plot considers all Ethernet drivers in Linux 4.19 and all network drivers in DPDK that do not rely on further external libraries.
We also omit DPDK drivers for FPGA boards (as these expect the user to bring their own hardware and driver support), unfinished drivers (Intel ice), drivers for virtual NICs, and 4 obsolete Linux drivers for which we could not identify the speed.
The linear relationship implies that driver complexity grows exponentially as network speed increases exponentially.

\subsection{Security Bugs in Operating Systems}
\label{sec:biscuitbugs}
Cutler et al. evaluate security bugs leading to arbitrary code execution in the Linux kernel found in 2017~\cite{biscuit}.
They identify 65 security bugs leading to code execution and find that 8 of them are use-after-free or double-free bugs, 32 are out-of-bounds accesses, 14 are logic bugs that cannot be prevented by the programming language, and 11 are bugs where the effect of the programming language remains unclear~\cite{biscuit}.
The 40 memory bugs (61\% of the 65 bugs) can be mitigated by using a memory-safe language such as their Go operating system Biscuit or our implementations. 
Performing an out-of-bounds access is still a bug in a memory-safe language, but it will only crash the program if it remains unhandled; effectively downgrading a code execution vulnerability to a denial of service vulnerability.
User space drivers can simply be restarted after a crash, crashing kernel drivers usually take down the whole system.

We analyze these 40 memory bugs identified by Cutler et al.~further and find that 39 of them are located in 11 different device drivers (the other bug is in the Bluetooth stack).
The driver with the most bugs in this data set is the Qualcomm WiFi driver with 13 bugs, i.e., a total of 20\% of all code execution vulnerabilities in the study could have been prevented if this network driver was written in a memory safe high-level language.
The key result here is that rewriting device drivers in safer languages achieves 97.5\% of the improvement they gained by rewriting the whole operating system.


\begin{table}[t]
 \setlength{\tabcolsep}{1.5mm}
	\centering
	\footnotesize
	\begin{tabular}{lrrrrrr}
		\textbf{Year} & \textbf{DPDK}$^*$ & \textbf{netmap}$^\dagger$ & \textbf{Snabb}$^*$ &  \textbf{PF\_RING}$^\dagger$ & \textbf{PSIO}$^\dagger$ & \textbf{PFQ}$^\dagger$ \\
		\toprule
		2010 & 0/0 & 0/0 & 0/0 & 0/0 & 1/1 & 0/0\\
		2011 & 0/0 & 0/0 & 0/0 & 1/1 & 0/0 & 0/0\\
		2012 & 0/1 & 1/4 & 0/0 & 0/4 & 0/1 & 0/1\\
		2013 & 0/0 & 3/8 & 0/0 & 0/0 & 0/1 & 0/0\\
		2014 & 4/11 & 4/14 & 0/0 & 0/4 & 1/1 & 1/2\\
		2015 & 9/15 & 6/14 & 1/2 & 4/7 & 2/3 & 1/2\\
		2016 & 12/22 & 1/12 & 0/1  & 0/1 & 0/0 & 1/1\\
		2017 & 17/23 & 3/10 & 0/0 & 1/2 & 1/1 & 1/1 \\
		\midrule
		\textbf{Sum} & 41/72 & 19/64 & 1/3 & 6/19 & 5/8 & 4/7\\
		\bottomrule
	\end{tabular}
        \begin{tablenotes}
		\item \footnotesize $^*$ User space driver
		\item \footnotesize $^\dagger$ Kernel driver with dedicated user space API
        \end{tablenotes}
        \vspace{1em}
	\caption{Packet processing frameworks used in academia, cells are uses/mentions; e.g., 1/3 means 3 papers mention the framework, 1 of them uses it}
	\label{tbl:paperstudy}
	\vspace{-3em}
\end{table}

\subsection{The Rise of DPDK}
\label{sec:risedpdk}
Section~\ref{sec:rw} used the open source projects Click, Open vSwitch, and pfSense as examples to show how network applications moved from running completely in the kernel to user space frameworks with a kernel driver (e.g., netmap) to full user space drivers (e.g., DPDK).
This trend towards DPDK is also present in academia.
We run a full-text search for all user space packet processing frameworks on all publications between 2010 and 2017 in the ACM conferences SIGCOMM and ANCS and the USENIX conferences NSDI, OSDI, and ATC ($n=1615$ papers).
This yields 113 papers which we skimmed to identify whether they merely mention a framework or whether they build their application on top of it or even present the framework itself.
Table~\ref{tbl:paperstudy} shows how DPDK started to dominate after it was open sourced in 2013\footnote{The paper mentioning DPDK in 2012 is the original netmap paper~\cite{netmap} referring to DPDK as a commercial offering with similar goals}.

\subsection{Languages~Used~for~DPDK~Applications}
Applications on top of DPDK are also not restricted to C, yet most users opt to use C when building a DPDK application.
This is a reasonable choice given that DPDK comes with C example code and C APIs, but there is no technical reason preventing programmers from using any other language.
In fact, DPDK applications in Rust, Go, and Lua exist~\cite{netbricks,gonff,moongen}.

DPDK's website showcases 21 projects building on DPDK. 14 (67\%) of these are written in C, 5 (24\%) in C++, 1 in Lua~\cite{moongen}, and 1 in Go~\cite{gonff}.
There are 116 projects using the \#dpdk topic on GitHub, 99 of these are applications using DPDK; the others are orchestration tools or helper scripts for DPDK setups.
69 (70\%) of these are written in C, 12 in C++, 4 in Rust (3 related to NetBricks~\cite{netbricks}, 1 wrapper), 4 in Lua (all related to MoonGen~\cite{moongen}), 3 in Go (all related to NFF-Go~\cite{gonff}), and one in Crystal (a wrapper).

Determining whether 67\% to 70\% of applications being written in C is unusually high requires a baseline to compare against.
Further up the network stack are web application frameworks (e.g., nodejs or jetty) that also need to provide high-speed primitives for applications on the next higher layer.
Of the 43 web platforms evaluated in the TechEmpower benchmark~\cite{techempowerbench} (the largest benchmark collection featuring all relevant web platforms) only 3 (7\%) are written in C.
17 different languages are used here, the dominant being Java with 19 (44\%) entries.
Of course not all platforms measured here are suitable for applications requiring high performance.
But even out of the 20 fastest benchmarked applications (``Single query'' benchmark) only one is written in C and 3 in C++. Java still dominates with 7 here.
Go (3), Rust (2), and C\# (1) are also present in the top 20.
This shows that it is possible to build fast applications on a higher layer in languages selected here.

\subsection{User Study: Mistakes in DPDK Applications Written in C}
\label{sec:userstudy}

We task students with implementing a simplified IPv4 router in C on top of DPDK for one of our networking classes.
The students are a mix of undergraduate and postgraduate students with prior programming experience, a basic networking class is a pre-requisite for the class.
Students are provided with a skeleton layer 2 forwarding program that handles DPDK setup, threading logic, and contains a dummy routing table.
Only packet validation according to RFC 1812~\cite{rfc1812}, forwarding via a provided dummy routing table implementation, and handling ARP is required for full credits.
ICMP and fragmentation handling is not required.

We received 55 submissions with at least partially working code, i.e., code that at least forwards some packets, in which we identified 3 types of common mistakes summarized in Table~\ref{tbl:userstudy}.
Incorrect programs can contain more than one class of mistake.
Logic errors are the most common and include failing to check EtherTypes, forgetting validations, and getting checksums wrong, these cannot be prevented by safer languages.
Memory bugs including use-after-free bugs can be prevented by the language.
No out-of-bounds access was made because the exercise offers no opportunity to do so: the code only needs to operate on the fixed-size headers which are smaller than the minimum packet size.
All integer overflow bugs happened due to code trying to decrement the time-to-live field in place in the packet struct without first checking if the field was already 0.

\emph{Ethical considerations.}
As handling errors done by humans requires special care we took all precautions to protect the privacy of the involved students.
The study was conducted by the original corrector of the exercise, these results are used for teaching the class.
No student code, answers, or any identifying information was ever given to anyone not involved in teaching the class. All submissions are pseudonymized.
We were able to achieve the ethical goals of avoiding correlating errors with persons.

\begin{table}[t]
 \setlength{\tabcolsep}{1.2mm}
	\centering
	\footnotesize
	\begin{tabular}{rrrrr}
		\textbf{Total} & \textbf{No mistakes} & \textbf{Logic error} & \textbf{Use-after-free} & \textbf{Int overflow} \\
		\toprule
		55 & 12 & 28 & 13 & 14 \\
		100\% & 22\% & 51\% & 24\% & 25\% \\
		\bottomrule
	\end{tabular}
	\caption{Mistakes made by students when implementing an IPv4 router in C on top of DPDK}
	\label{tbl:userstudy}
	\vspace{-3em}
\end{table}

\subsection{Summary}
To summarize: (network) drivers are written in C which leads to preventable bugs in both the drivers (Sec. \ref{sec:biscuitbugs}) and in applications on top of them (Sec.~\ref{sec:userstudy}).
Historical reasons for writing them in C no longer apply (Sec.~\ref{sec:risedpdk}).
Driver complexity is growing (Sec.~\ref{sec:driversloc}), so let's start using safer languages.

\section{Implementations}
\label{sec:impl}
All of our implementations are written from scratch by experienced programmers in idiomatic style for the respective language.
We target Linux on x86 using the \texttt{uio} subsystem to map PCIe resources into user space programs.
For details on implementing user space drivers, have a look at our original publication about the C driver~\cite{ixy}.

\subsection{Architecture}
All of our drivers implement the same basic architecture as our ixy user space driver~\cite{ixy} which serves as reference implementation that our high-level drivers are compared against.
Ixy's architecture is inspired by DPDK, i.e., it uses poll mode drivers without interrupts, all APIs are based on batches of packets, and DMA buffers are managed in custom memory pools.
The memory pool for DMA buffers is also needed despite automatic memory management: not all memory is suitable for use as DMA buffers, simply using the language's allocator is not possible.
See~\cite{ixy} for details on DMA allocation for user space drivers.
This restriction potentially circumvents some of the memory safety properties of the languages in parts of the driver.

 It is a common misconception that user space drivers do not support interrupts in general.
The Linux \texttt{vfio} subsystem offers full support for interrupts in user space drivers via an event file descriptor allowing handling interrupts with normal IO syscalls~\cite{vfio} that are available in high-level languages.
However, interrupts lead to packet loss for packet rates above 2\,Mpps on our test systems in a C implementation.
They are useful as power-saving mechanism under low load and should be supported in a real-world driver, but we did not implement support for them in most drivers as we are mainly interested in peak performance.

\subsection{Challenges for High-Level Languages}
There are three main challenges for user space drivers in high-level languages compared to kernel drivers in C.

\subsubsection{Handling external memory}
Two memory areas cannot be managed by the language itself: Memory-mapped IO regions are provided by the device and DMA buffers must be allocated with a custom allocator.
Languages need access to the \texttt{mmap} and \texttt{mlock} syscalls to allocate these memory areas.
We use a small C function in languages where these syscalls are either not available at all or only supported with restricted flags.

\subsubsection{Unsafe primitives}
External memory, i.e., PCIe address space and DMA buffers, must be wrapped in language-specific constructs that enforce bounds checks and allow access to the backing memory.
Many languages come with dedicated wrapper structs that are constructed from a raw pointer and a length.
For other languages we have to implement these wrappers ourselves.

In any case, all drivers need to perform inherently unsafe memory operations that cannot be checked by any language feature.
The goal is to restrict these unsafe operations to as few places as possible to reduce the amount of code that needs to be manually audited for memory errors.

\subsubsection{Memory access semantics}
Memory-mapped IO regions are memory addresses that are not backed by memory, each access is forwarded to a device and handled there.
Simply telling the language to read or write a memory location in these regions can cause problems as optimizers make assumptions about the behavior of the memory.
For example, writing a control register and never reading it back looks like a dead store to the language and the optimizer is free to remove the access.
Repeatedly reading the same register in a loop while waiting for a value to be changed by the device looks like an opportunity to hoist the read out of the loop.

C solves this problem with the \texttt{volatile} keyword guaranteeing that at least one read or write access is performed.
The high-level language needs to offer control over how these memory accesses are performed.
Atomic accesses and memory barriers found in concurrency utilities make stronger guarantees and can be used instead if the language does not offer volatile semantics.
Primitives from concurrency utilities can also substitute the access-exactly-once semantics required for some device drivers. 

Readers interested in gory details about memory semantics for device drivers are referred to the Linux kernel documentation on memory barriers~\cite{kernel-memory-barriers}.
It is worth noting that all modern CPU architectures offer a memory model with cache-coherent DMA simplifying memory handling in drivers.
We only test on x86 as proof of concept, but DPDK's support for x86, ARM, POWER, and Tilera shows that user space drivers themselves are possible on a wide range of modern architectures.
Some of our implementations in dynamic languages likely rely on the strong memory model of x86 and might require modifications to work reliably on architectures with a weaker memory model such as ARM.

\subsection{Rust Implementation}
\label{sec:rust}
Rust is an obvious choice for safe user space drivers: safety and low-level features are two of its main selling points.
Its ownership-based memory management allows us to use the native memory management even for our custom DMA buffers.
We allocate a lightweight Rust struct for each packet that contains metadata and owns the raw memory.
This struct is essentially being used as a smart pointer, i.e., it is often stack-allocated.
This object is in turn either owned by the memory pool itself, the driver, or the user's application.
The compiler enforces that the object has exactly one owner and that only the owner can access the object, this prevents use-after-free bugs despite using a completely custom allocator.
Rust is the only language evaluated here that protects against use-after-free bugs and data races in memory buffers.

External memory is wrapped in \texttt{std::slice} objects that enforce bounds checks on each access, leaving only one place tagged as unsafe that can be the source of memory errors: we need to pass the correct length when creating the slice.
Volatile memory semantics for accessing device registers are available in the \texttt{ptr} module.

\subsubsection{IOMMU and interrupt support}
We also implemented support for \texttt{vfio} in the Rust driver, all other implementations only support the simpler \texttt{uio} interface.
This interface enables us to use the IOMMU to isolate the device and run without root privileges and to use interrupts instead of polling under low load.
Moreover, this interface offers a portable approach to allocating DMA memory in user space; the other implementations are specific to x86 and rely on an implementation detail in Linux (see~\cite{ixy}).

\subsubsection{Related work}
NetBricks (2016)~\cite{netbricks} is a network function framework that allows users to build and compose network functions written in Rust.
It builds on DPDK, i.e., the drivers it uses are written in C.
They measure a performance penalty of 2\% to 20\% for Rust vs. C depending on the network function being executed.

We also ported our driver to Redox (2015)~\cite{redox}, a real-world operating system under active development featuring a microkernel written in Rust with user space drivers.
It features two network drivers for the Intel e1000 NIC family (predecessor of the ixgbe family used here) and Realtek rtl8168 NICs.
Table~\ref{tbl:rust-unsafe} compares how much unsafe code their drivers use compared to our implementations.
Inspecting the pre-existing Redox drivers shows several places where unsafe code could be made safe with some more work as showcased by our Redox port.
Line count for our Linux driver includes all logic to make user space drivers on Linux work, our Linux version therefore has more unsafe code than the Redox version which already comes with the necessary functionality.
These line counts also show the relationship between NIC speed and driver complexity hold true even for minimal drivers in other systems.

\begin{table}[t]
 \setlength{\tabcolsep}{1.2mm}
	\centering
	\footnotesize
	\begin{tabular}{lrrrr}
		\textbf{} & \textbf{} & \textbf{Code} & \textbf{Unsafe} & \textbf{} \\
		\textbf{Driver} & \textbf{NIC Speed} & \textbf{[Lines]} & \textbf{[Lines]} & \textbf{\% Unsafe} \\
		\toprule
		Our Linux implementation & 10\,Gbit/s & 961 & 125 & 13.0\% \\
		Our Redox implementation & 10\,Gbit/s & 901 & 68 & 7.5\% \\
		Redox e1000 & 1\,Gbit/s & 393 & 140 & 35.6\% \\
		Redox rtl8168 & 1\,Gbit/s & 363 & 144 & 39.8\% \\
		\bottomrule
	\end{tabular}
	\caption{Unsafe code in different Rust drivers}
	\label{tbl:rust-unsafe}
	\vspace{-3em}
\end{table}

\subsection{Go Implementation}
Go is a compiled systems programming language maintained by Google that is often used for distributed systems.
Memory is managed by a garbage collector tuned for low latency.
It achieved pause times in the low millisecond range in 2015~\cite{go-fast-gc} and sub-millisecond pause times since 2016~\cite{go-faster-gc}.

External memory is wrapped in slices to provide bounds checks.
Memory barriers and volatile semantics are indirectly provided by the atomic package which offers primitives with stronger guarantees than required.

\subsubsection{Related work}
Biscuit (2018)~\cite{biscuit} is a research operating system written in Go that features a network driver for the same hardware as we are targeting here.
Unlike all other research operating systems referenced here, they provide an explicit performance comparison with C.
They observe GC pauses of up to 115\,$\mu$s in their benchmarks and an overall performance of 85\% to 95\% of an equivalent C version.
Unfortunately it does not offer a feasible way to benchmark only the driver in isolation for a comparison. 

NFF-GO (2017)~\cite{gonff} is a network function framework allowing users to build and compose network functions in Go.
It builds on DPDK, i.e., the drivers it uses are written in C.
Google's Fuchsia~\cite{fuchsia} mobile operating system features a TCP stack written in Go on top of C drivers.

\subsection{C\# Implementation}
C\# is a versatile JIT-compiled and garbage-collected language offering both high-level features and systems programming features.
Several methods for handling external memory are available, we implemented support for two of them to compare them.
\texttt{Marshal} in \texttt{System.Runtime.InteropServices} offers wrappers and bounds-checked accessors for external memory.
C\# also offers a more direct way to work with raw memory: its unsafe mode enables language features similar to C, i.e., full support for pointers with no bounds checks and volatile memory access semantics.


\subsubsection{Related work}
The Singularity (2004) research operating system~\cite{singularity} is written in Sing\#, a dialect of C\# with added contracts and safety features developed for use in Singularity.
It comes with a driver for Intel 8254x PCI NICs that are architecturally similar to the 82599 NICs used here: their DMA ring buffers are virtually identical.
All memory accesses in their drivers are facilitated by safe APIs offered by the Singularity kernel using contracts to ensure that the driver cannot access memory it is not supposed to access.

\subsection{Java Implementation}
\label{sec:java}
Java is a JIT-compiled and garbage-collected language similar to C\# (which was heavily inspired by Java).
The only standardized way to access external memory is by calling into C using JNI, a verbose and slow foreign function interface.
We target OpenJDK 12 which offers a non-standard way to handle external memory via the \texttt{sun.misc.Unsafe} object that provides functions to read and write memory with volatile access semantics.
We implement and compare both methods here.
Java's low-level features are inferior compared to C\#, the non-standard \texttt{Unsafe} object is cumbersome to use compared to C\#'s unsafe mode with full pointer support.
Moreover, Java does not support unsigned integer primitives requiring work-arounds as hardware often uses such types.

\subsubsection{Related work}
JavaOS (1996)~\cite{javaos} was a commercial operating system targeting embedded systems and thin clients written in Java, it was discontinued in 1999.
Their device driver guide~\cite{javaos-drivers} contains the source code of a 100\,Mbit/s network driver written in Java as an example.
The driver implements an interface for network drivers and calls out to helper functions and wrapper provided by JavaOS for all low-level memory operations.

\subsection{OCaml Implementation}
OCaml is a compiled functional language with garbage collection.
We use OCaml \texttt{Bigarray}s backed by external memory for DMA buffers and PCIe resources, allocation is done via C helper functions.
The \texttt{Cstruct} library~\cite{ocaml-cstruct} from the MirageOS project~\cite{mirage} allows us to access data in the arrays in a structured way by parsing definitions similar to C struct definitions and generating code for the necessary accessor functions.

\subsubsection{Related work}
We also ported our driver to MirageOS (2013)~\cite{mirage}, a unikernel written in OCaml with the main goal of improving security.
MirageOS is not optimized for performance (e.g., all packets are copied when being passed between driver and network stack) and no performance evaluation is given by its authors (performance regression tests are being worked on~\cite{mirage-perf-tests}).
MirageOS targets Xen, KVM, and normal Unix processes. The Xen version has a driver-like interface for Xen netfront (not a PCIe driver, though), the KVM version builds on the Solo5 unikernel execution environment~\cite{solo5} that provides a VirtIO driver written in C.
Our port is the first PCIe network driver written in OCaml in MirageOS, currently targeting mirage-unix as the other versions lack the necessary PCIe support.

\subsection{Haskell Implementation}
Haskell is a compiled functional language with garbage collection.
All necessary low-level memory access functions are available via the \texttt{Foreign} package.
Memory allocation and mapping is available via \texttt{System.Posix.Memory}.

\subsubsection{Related work}
House (2005)~\cite{house} is a research operating system written in Haskell focusing on safety and formal verification using P-Logic~\cite{plogic}.
It provides a monadic interface to memory management, hardware, user-mode processes, and low-level device IO.
No quantitative performance evaluation is given.

PFQ~\cite{pfq} is a packet processing framework offering a Haskell interface and pfq-lang, a specialized domain-specific language for packet processing in Haskell.
It runs on top of a kernel driver in C.
Despite the focus on Haskell it is mainly written in C as it relies on C kernel modules: PFQ is 75\% C, 10\% C++, 7\% Haskell by lines of code.

\subsection{Swift Implementation}
Swift is a compiled language maintained by Apple mainly targeted at client-side application development.
Memory in Swift is managed via automatic reference counting, i.e., the runtime keeps a reference count for each object and frees the object once it is no longer in use. 
Despite primarily targeting end-user applications, Swift also offers all features necessary to implement drivers.
Memory is wrapped in \texttt{UnsafeBufferPointer}s (and related classes) that are constructed from an address and a size.
Swift only performs bounds checks in debug mode.

\subsubsection{Related work}
No other drivers or operating systems in Swift exist.
The lowest level Swift projects that are available are the Vapor~\cite{vapor} and Kitura~\cite{kitura} frameworks for server-side Swift.

\subsection{JavaScript Implementation}
\label{sec:javascript}
We build on Node.js~\cite{node} with the V8 JIT compiler and garbage collector, a common choice for server-side JavaScript.
We use \texttt{ArrayBuffer}s to wrap external memory in a safe way, these arrays can then be accessed as different integer types using \texttt{TypedArray}s, circumventing JavaScript's restriction to floating point numbers.
We also use the \texttt{BigInt} type that is not yet standardized but already available in Node.js.
Memory allocation and physical address translation is handled via a Node.js module in C.

\subsubsection{Related work}
JavaScript is rarely used for low-level code, the most OS-like projects are NodeOS~\cite{nodeos-github} and OS.js~\cite{osjs}.
NodeOS uses the Linux kernel with Node.js as user space.
OS.js runs a window manager and applications in the browser and is backed by a server running Node.js on a normal OS.
Neither of these implements driver-level code in JavaScript.

\subsection{Python Implementation}
\label{sec:python}
Python is an interpreted scripting language with garbage collection.
Our implementation uses Cython for handling memory (77 lines of code), the remainder of the driver is written in pure Python.
Performance is not the primary goal of this version of our driver, it is the only implementation presented here that is not explicitly optimized for performance.
It is meant as a simple prototyping environment for PCIe drivers and as an educational tool.

Writing drivers in scripting languages allows for quick turn-around times during development or even an explorative approach to understanding hardware devices in an interactive shell.
We provide primitives for PCIe driver development in Python that we hope to be helpful to others as this is the first PCIe driver in Python to our knowledge.

\subsubsection{VirtIO driver}
We also implemented a driver for virtual VirtIO~\cite{virtiospec} NICs here to make this driver accessible to users without dedicated hardware.
A provided Vagrant~\cite{vagrant} file allows spinning up a test VM to get started with PCIe driver development in Python in a safe environment.

\subsubsection{Related work}
Python is a popular~\cite{pyusb-stats} choice for user space USB drivers with the PyUSB library~\cite{pyusb}. 
In contrast to our driver, it is mainly a wrapper for a C library.
Python USB drivers are used for devices that either mainly rely on bulk transfers (handled by the underlying C library) or that do not require many transfers per second. 

\section{Evaluation}
\label{sec:eval}

\begin{table}[t]
 \setlength{\tabcolsep}{1.1mm}
	\centering
	\footnotesize
	\begin{tabular}{lrrr}
		\\
		\textbf{Lang.} & \textbf{Lines of code}$^1$ & \textbf{Lines of C code}$^1$  & \textbf{Source size (gzip$^2$)} \\
		\toprule
		C~\cite{ixy} & 831 & 831 & 12.9\,kB  \\
		Rust & 961 & 0 & 10.4\,kB\\
		Go & 1640 & 0 & 20.6\,kB \\
		C\# & 1266 & 34 & 13.1\,kB\\
		Java & 2885 & 188 & 31.8\,kB \\
		OCaml & 1177 & 28 &  12.3\,kB\\
		Haskell & 1001 & 0 &  9.6\,kB\\
		Swift & 1506 & 0 & 15.9\,kB \\
		JavaScript & 1004 & 262 &13.0\,kB \\
		Python & 1242 & (Cython) 77 &14.2\,kB \\
		\bottomrule
	\end{tabular}
	\begin{tablenotes}
	\item $^1$ Incl. C code, excluding empty lines and comments, counted with \texttt{cloc}
	\item $^2$ Compression level 6
	\end{tablenotes}
	\caption{Size of our implementations stripped down to the core feature set}
	\label{tbl:lang-lines}
	\vspace{-3em}
\end{table}

Table~\ref{tbl:lang-lines} compares the code size as counted with \texttt{cloc} ignoring empty lines and comments, we also include the code size after compressing it with gzip to estimate information entropy as lines of code comparisons between different languages are not necessarily fair.
We stripped features not present in all drivers (i.e., all (unit-)tests, alternate memory access implementations, VirtIO support in C and Python, IOMMU/interrupt support in C and Rust) for this evaluation.
We also omit register definitions because several implementations contain automatically generated lists of $> 1000$ mostly unused constants for register offsets.
All high-level languages require more lines than C, but the Rust, Haskell, and OCaml implementations are smaller in size as their formatting style leads to many short lines.
Java and JavaScript require more C code due to boilerplate requirements of their foreign function interfaces.

Table~\ref{tbl:lang-safety} summarizes protections against classes of bugs available to both our driver implementations and applications built on top of them.
The take-away here is that high-level languages do not necessarily increase the work-load for the implementor while gaining safety benefits.
Subjectively, we have even found it easier to write driver code in high-level languages --- even if more lines of code were required --- after figuring out the necessary low-level details of the respective language (a one-time effort).
We also discovered memory bugs in the original C implementation while porting the drivers.

\begin{table}[t]
 \setlength{\tabcolsep}{0.8mm}
	\centering
	\footnotesize
	\begin{tabular}{lccccc}
		& \multicolumn{2}{c}{\textbf{General memory}} & \multicolumn{2}{c}{\hspace{-1em}\textbf{Packet buffers}}  \\
		\textbf{Lang.} & \textbf{OoB$^1$} & \textbf{Use after free}  & \textbf{OoB$^1$} & \textbf{Use after free} & \textbf{Int overflows} \\
		\toprule
		C~\cite{ixy} & \xmark & \xmark & \xmark & \xmark & \xmark \\
		Rust & \cmark & \cmark & (\cmark)$^2$ & \cmark & (\cmark)$^5$ \\
		Go & \cmark & \cmark & (\cmark)$^2$ & (\cmark)$^4$ & \xmark \\
		C\# & \cmark & \cmark & (\cmark)$^2$ & (\cmark)$^4$ & (\cmark)$^5$ \\
		Java & \cmark & \cmark & (\cmark)$^2$ & (\cmark)$^4$ & \xmark \\
		OCaml & \cmark & \cmark & (\cmark)$^2$ & (\cmark)$^4$ & \xmark \\
		Haskell & \cmark & \cmark & (\cmark)$^2$ & (\cmark)$^4$ & (\cmark)$^6$ \\
		Swift & \cmark & \cmark & \xmark$^3$ & (\cmark)$^4$ & \cmark \\
		JavaScript & \cmark & \cmark & (\cmark)$^2$ & (\cmark)$^4$ & (\cmark)$^6$ \\
		Python & \cmark & \cmark & (\cmark)$^2$ & (\cmark)$^4$ & (\cmark)$^6$ \\
		\bottomrule
	\end{tabular}
	\begin{tablenotes}
	\item $^1$ Out of bounds accesses
	\item $^2$ Bounds enforced by wrapper, constructor in unsafe or C code
	\item $^3$ Bounds only enforced in debug mode
	\item $^4$ Buffers are never free'd/gc'd, only returned to a memory pool
	\item $^5$ Disabled by default
	\item $^6$ Uses floating point or arbitrary precision integers by default
	\end{tablenotes}
	\caption{Language-level protections against classes of bugs in our drivers and the C reference code}
	\label{tbl:lang-safety}
	\vspace{-3em}
\end{table}

\section{Performance}
\label{sec:tp}

The limiting factor for network drivers is the number of packets per second, not the bandwidth.
All tests therefore use minimum-sized packets (up to 29.76\,Mpps at 20\,Gbit/s).
We also benchmark the reference C driver ixy~\cite{ixy} as baseline performance measurement.
The C driver proved to be as fast as older versions of DPDK but slower than modern versions of DPDK that offer a heavily vectorized transmit and receive path in its ixgbe driver~\cite{ixy}.

\subsection{Test Setup} We run our drivers on a Xeon E3-1230 v2 CPU clocked at 3.3\,GHz with two 10\,Gbit/s Intel X520 NICs.
Test traffic is generated with MoonGen~\cite{moongen}.
All drivers configure the NICs with a queue size of 512 (evaluation of different queue sizes can be found in \cite{ixy}) and run a forwarding application that modifies one byte in the packet headers.

All tests are restricted to a single CPU core even though most of our implementations feature multi-core support.
We already hit hardware limits with only one core.
Multi-core scaling for network applications can be done at the hardware level even if the language does not support multi-threading.
The NIC can split up incoming packets by hashing the flow and deliver them to independent processes.
This enables trivial multi-core scaling independent of language restrictions, e.g., Snabb~\cite{snabb} uses multiple processes to scale the single-threaded LuaJIT runtime to multiple cores.

\begin{figure*}[t!]
\centering
    \subfigure[Forwarding rate on 1.6\,GHz CPU]{
        \includegraphics[width=0.47\textwidth]{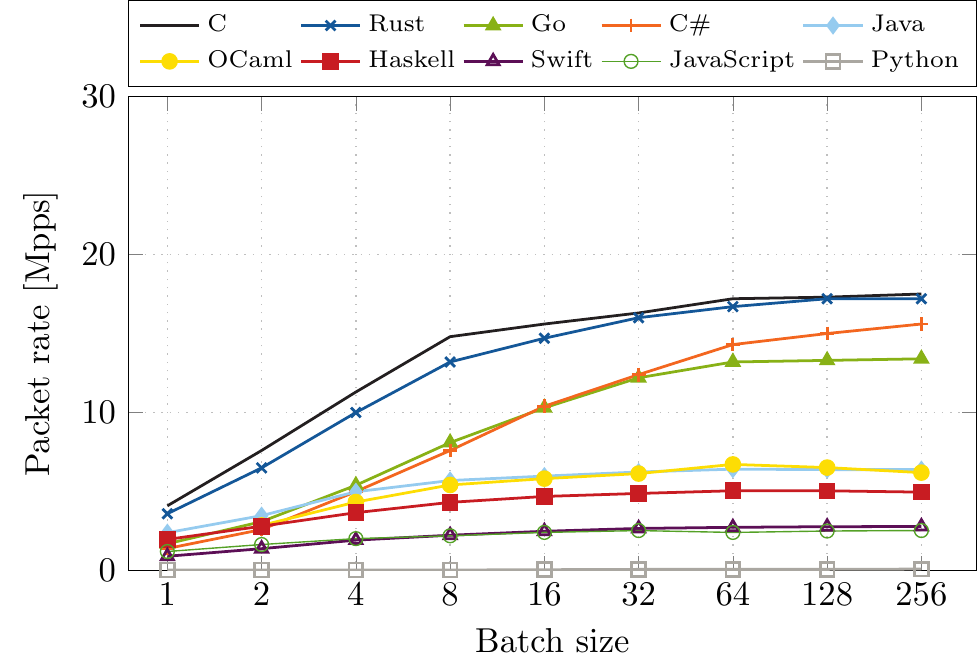}
        \label{fig:batches16}
    }
    \subfigure[Forwarding rate on 3.3\,GHz CPU]{
        \includegraphics[width=0.47\textwidth]{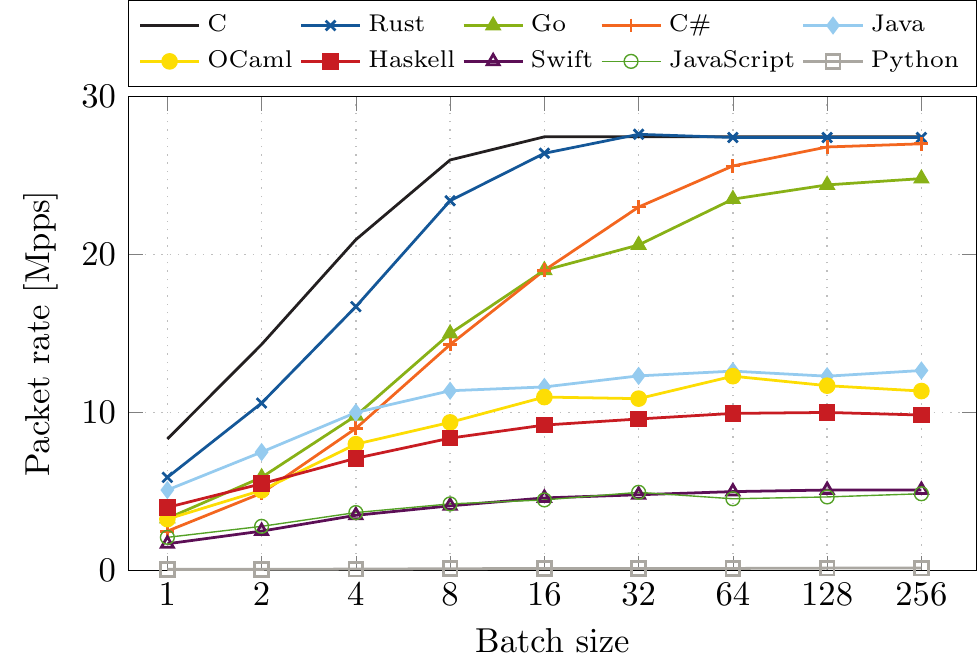}
        \label{fig:batches33}
    }
    \vspace{-0.75em}
    \caption{Forwarding rate of our implementations with different batch sizes}
    \label{fig:batches}
\end{figure*}

\subsection{Effect of Batch Sizes}

Processing packets in batches is a core concept of all fast network drivers~\cite{gallenmuancs}.
Each received or sent batch requires synchronization with the network card, larger batches therefore improve performance.
Too large batch sizes fill up the CPU's caches so there are diminishing returns or even reduced performance.
32 to 128 packets per batch is the sweet spot for user space drivers~\cite{gallenmuancs,batchingclick,vppperf}.

Figure~\ref{fig:batches} shows the maximum achievable bidirectional forwarding performance of our implementations.
We also run the benchmark at a reduced CPU frequency of 1.6\,GHz as the performance of the C and Rust forwarders quickly hit some hardware bottleneck at 94\% line rate at 3.3\,GHz.
A few trade-offs for the conflicting goals of writing idiomatic safe code and achieving a high-performance were evaluated for each language.
Haskell and OCaml allocate a new list/array for each processed batch of packets while all other languages re-use arrays.
Recycling arrays in these functional languages building on immutable data structures would not be idiomatic code, this is one of the reasons for their lower performance.

\subsubsection{Rust}
Rust achieves 90\% to 98\% of the baseline C performance.
Our Redox port of the Rust driver only achieves 0.12\,Mpps due to performance bottlenecks in Redox (high performance networking is not yet a goal of Redox).
This is still a factor 150 improvement over the pre-existing Redox drivers due to support for asynchronous packet transmission.

\subsubsection{Go}
Go also fares well, proving that it is a suitable candidate for a systems programming language.

\subsubsection{C\#}
The aforementioned utility functions from the \texttt{Marshal} class to handle memory proved to be too slow to achieve competitive performance.
Rewriting the driver to use C\# unsafe blocks and raw pointers in selected places improved performance by 40\%.
Synthetic benchmarks show a 50-60\% overhead for the safer alternatives over raw pointers.

\subsubsection{Java}\label{sec:java-perf}
Our implementation heavily relies on devirtualization, inlining, and dead-code elimination by the optimizer as it features several abstraction layers and indirections that are typical for idiomatic Java code.
We used profiling to validate that all relevant optimization steps are performed, almost all CPU time is spent in the transmit and receive functions showing up as leaf functions despite calling into an abstraction layer for memory access.

We use OpenJDK 12 with the HotSpot JIT, the \emph{Parallel} garbage collector, and the Unsafe object for memory access.
Using JNI for memory access instead is several orders of magnitude slower.
Using OpenJ9 as JIT reduces performance by 14\%.
These results are significantly slower than C\# and show that C\#'s low-level features are crucial for fast drivers.
One reason is that C\# features \emph{value types} that avoid unnecessary heap allocations.
We try to avoid object allocations by recycling packet buffers, but writing allocation-free code in idiomatic Java is virtually impossible, so there are some allocations.
OpenJDK 12 features 7 different garbage collectors which have an impact on performance as shown in Table~\ref{tbl:java-gc-perf}.
Epsilon is a no-op implementation that never frees memory, leaking approximately 20\,bytes per forwarded packet.

\begin{table}[b]
	 \setlength{\tabcolsep}{0.95mm}
	\centering
	\footnotesize
	\begin{tabular}{rrrrrrrr}
		\textbf{Batch\textbackslash{}GC} & \textbf{CMS} & \textbf{Serial} & \textbf{Parallel} & \textbf{G1} & \textbf{ZGC} & \textbf{Shenandoah} & \textbf{Epsilon} \\
		\toprule
		\textbf{4} & 9.8 & 10.0 & 10.0 & 7.8 & 9.4 & 8.5 & 10.0 \\
		\textbf{32} & 12.3 & 12.4 & 12.3 & 9.3 & 11.5 & 10.8 & 12.2 \\
		\textbf{256} & 12.6 & 12.4 & 12.3 & 9.7 & 11.6 & 11.2 & 12.7 \\
		\bottomrule
	\end{tabular}
	\caption{Performance of different Java garbage collectors in Mpps when forwarding packets at 3.3\,GHz}
	\label{tbl:java-gc-perf}
\end{table}

\subsubsection{OCaml}
Enabling the Flambda~\cite{flambda} optimizations in OCaml 4.07.0 increases throughput by 9\%.
An interesting optimization is representing bit fields as separate 16 bit integers instead of 32 bit integers if possible: Larger integers are boxed in the OCaml runtime.
This increases performance by 0.7\% when applied to the status bits in the DMA descriptors.

Our MirageOS port achieves 3\,Gbit/s TCP throughput using iperf and the Mirage TCP stack.
The bottleneck is MirageOS as it lacks support for batching and requires several copy operations for each packet.

\subsubsection{Haskell}
Compiler (GHC 8.4.3) optimizations seem to do more harm than good in this workload.
Increasing the optimization level in the default GHC backend from \texttt{O1} to \texttt{O2} reduces throughput by 11\%.
The data in the graph is based on the LLVM backend which is 3.5\% faster than the default backend at \texttt{O1}.
Enabling the threaded runtime in GHC decreases performance by 8\% and causes the driver to lose packets even at loads below 1\,Mpps due to regular GC pauses of several milliseconds.

\subsubsection{Swift}
Swift increments a reference counter for each object passed into a function and decreases it when leaving the function.
This is done for every single packet as they are wrapped in Swift-native wrappers for bounds checks.
There is no good way to disable this behavior for the wrapper objects while maintaining an idiomatic API for applications using the driver.
A total of 76\% of the CPU time is spent incrementing and decrementing reference counters.
This is the only language runtime evaluated here that incurs a large cost even for objects that are never free'd.

\subsubsection{JavaScript}
We also compare different Node.js versions: 10 (V8 6.9, current LTS), 11 (V8 7.0), and 12 (V8 7.5), older versions are unsupported due to lack of \texttt{BigInt} support.
Node 10 and 11 perform virtually identical, upgrading to 12 degrades performance by 13\% as access to \texttt{TypedArray}s is slower in this version.
Performance optimizations applied to reach the current level are related to access \texttt{TypedArray}s (which are faster than plain \texttt{DataView}s) and reducing usage of \texttt{BigInt}s: especially constructing BigInts is slow and most values required can be represented in the double data type.

\subsubsection{Python}
Python only achieves 0.14\,Mpps in the best case using the default CPython interpreter in version 3.7.
Most time is spent in code related to making C structs available to higher layers.
We are using constructs that are incompatible with the JIT compiler PyPy.
This is the only implementation here not optimized for performance, we believe it is possible to increase throughput by an order of magnitude by re-cycling struct definitions.
Despite this we are content with the Python implementation as the main goal was not a high throughput but a simple implementation in a scripting language for prototyping functionality.

\begin{table}[t]
 \setlength{\tabcolsep}{1.1mm}
	\centering
	\footnotesize
	\begin{tabular}{lrrrrrrrr}
		& \multicolumn{4}{c}{Batch 32, 1.6\,GHz} & \multicolumn{4}{c}{Batch 8, 1.6\,GHz} \\
		\textbf{Events per packet} & \hspace{1em} & \textbf{C} & \textbf{Rust} & \hspace{2.5em} & & \textbf{C} & \textbf{Rust} & \hspace{1.5em} \\
		\toprule
		\textbf{Cycles}                     & & 94 & 100     &&& 108 &  120 \\
		\textbf{Instructions}              & & 127 & 209   &&& 139 &  232  \\
		\textbf{Instr. per cycle}         & & 1.35 & 2.09 &&& 1.29 & 1.93 &  \vspace{0.35em}  \\
		\textbf{Branches}                 & & 18 & 24      &&& 19 &  27  \\
		\textbf{Branch mispredicts} & & 0.05 & 0.08       &&& 0.02 & 0.06 & 		\vspace{0.35em} \\
		\textbf{Store $\mu$ops}       & & 21.8 & 37.4      &&& 24.4 & 43.0  \\
		\textbf{Load $\mu$ops}       & & 30.1 & 77.0      &&& 33.4 & 84.2\  \\
		\textbf{Load L1 hits}                     & & 24.3 & 75.9      &&& 28.8 & 83.1 \\
		\textbf{Load L2 hits}                     & & 1.1 & 0.05         &&& 1.2 & 0.1 \\
		\textbf{Load L3 hits}                     & & 0.9 & 0.0      &&& 0.5 & 0.0 \\
		\textbf{Load L3 misses}               & & 0.3 & 0.1         &&& 0.3 & 0.3 \\
		\bottomrule
	\end{tabular}
	\caption{Performance counter readings in events per packet when forwarding packets}
	\label{tbl:rust-profiling}
	\vspace{-3em}
\end{table}
\subsection{The Cost of Safety Features in Rust}
Rust is our fastest implementation achieving more than 90\% of the performance of C when constrained by available CPU resources.
It is also the only high-level language without overhead for memory management here, making it an ideal candidate to investigate overhead further by profiling.
There are only two major differences between the Rust and C implementations:
\begin{itemize}
\item[(1)] Rust enforces bounds checks while the C code contains no bounds checks (arguably idiomatic style for C).
\item[(2)] C does not require a wrapper object for DMA buffers, it stores all necessary metadata directly in front of the packet data the same memory area as the DMA buffer.
\end{itemize}
However, the Rust wrapper objects can be stack-allocated and effectively replace the pointer used by C with a smart pointer, mitigating the locality penalty.
The main performance disadvantage is therefore bounds checking.

We use CPU performance counters to profile our forwarder with two different batch sizes.
Table~\ref{tbl:rust-profiling} lists the results in events per forwarded packet.
Rust requires 65\% (67\%) more instructions to forward a single packet at a batch size of 32 (8).
The number of branches executed rises by 33\% (42\%), the number of loads even by 150\% (180\%).
However, the Rust code only requires 6\% (11\%) more cycles per packet overall despite doing more work.
Synthetic benchmarks can achieve an even lower overhead of bounds checking~\cite{flater2013case}.
A modern superscalar out of order processor can effectively hide the overhead introduced by these safety checks: normal execution does not trigger bounds check violations, the processor is therefore able to correctly predict (branch mispredict rate is at 0.2\% - 0.3\%) and speculatively execute the correct path.\footnote{A good example of speculatively executing code in the presence of bounds checks is the Spectre v1 security vulnerability which exists due to this performance optimization in CPUs~\cite{spectrev1}. Note that user space drivers are not affected by this vulnerability as the control flow does not cross a trust boundary as everything runs in the same process.}
The Rust code achieves about 2 instructions per cycle vs. about 1.3 instructions with the C code.

Caches also help with the additional required loads of bounds information: this workload achieves an L1 cache hit rate of 98.5\% (98.7\%).
Note that the sum of cache hits and L3 misses is not equal to the number of load $\mu$ops because some loads are executed immediately after the store, fetching the data from the store buffer before it even reaches the cache.

Another safety feature in Rust are integer overflow checks that catch a common mistake in our user study, see Section~\ref{sec:userstudy}.
Overflow checks are currently disabled by default in release mode in Rust and have to be explicitly enabled with a compile-time flag.
Doing so decreases throughput by only 0.8\% at batch size 8, no statistically significant deviation was measurable with larger batch sizes.
Profiling shows that 9 additional instructions per packet are executed by the CPU, 8 of them are branches.
Total branch mispredictions are unaffected, i.e., the  branch check is always predicted correctly by the CPU.
This is another instance of speculative execution in an out-of-order CPU hiding the cost of safety features.

\begin{table}[b]
 \setlength{\tabcolsep}{0.6mm}
	\centering
	\footnotesize
	\begin{tabular}{lrrrrrrrrr}
		\textbf{Bench.\textbackslash{}Lang.} & \textbf{Rust} & \textbf{Go} & \textbf{C\#} & \textbf{Java} & \textbf{OCaml} & \textbf{Haskell} & \textbf{Swift} &\textbf{JS} & \textbf{Py.} \\
		\toprule
		\textbf{Our results}  & 98\%& 81\%& 76\%& 38\% & 38\%& 30\%& 16\%& 16\%& 1\%\\
		\textbf{CLBG}~\cite{lang-benchmarks} & 117\%& 34\%& 73\%& 52\% & 80\%& 65\%& 64\%& 28\%& 4\%\\
		\bottomrule
	\end{tabular}
	\caption{Performance results normalized to C, i.e., 50\% means it achieves half the speed of C}
	\label{tbl:lang-benchmark}
	\vspace{-3em}
\end{table}
\subsection{Comparison with Other Language Benchmarks}
Table~\ref{tbl:lang-benchmark} compares our performance results with the Computer Language Benchmarks Game (CLBG)~\cite{lang-benchmarks}, a popular more general performance comparison of different languages.
We use the ``fastest measurement at the largest workload'' data set from 2019-07-21, summarized as geometric mean~\cite{fleming1986not} over all benchmarks.
Our results are for batch size 32 (realistic value for real-world applications, e.g., DPDK default) at 1.6\,GHz CPU speed to enforce a CPU bottleneck.

This shows that especially dynamic and functional languages pay a performance penalty when being used for low-level code compared to the more general benchmark results.
Our implementations (except Python) went through several rounds of optimization based on profiling results and extensive benchmarks.
While there are certainly some missed opportunities for further minor optimization, we believe to be close to the optimum achievable performance for drivers in idiomatic code in these languages.
Note that the reference benchmark is also probably not perfect.
Go and C\# perform even better at the low-level task of writing drivers than the general purpose benchmarks, showing how a language's performance characteristics depend on the use case.
General-purpose benchmark results can be misleading when writing low-level code in high-level languages.

\section{Latency}
\label{sec:latency}
Latency is dominated by time spent in buffers, not by time spent handling a packet on the CPU.
Our drivers forward packets in hundreds of cycles, i.e., within hundreds of nanoseconds.
A driver with a lower throughput is therefore not automatically one with a higher latency while operating below its load limit.
The main factors driving up latency are pauses due to garbage collection and the batch size.
Note that the batch size parameter is only the maximum batch size, a driver operating below its limit will process smaller batches.
Drivers running closer to their limits will handle larger batches and incur a larger latency.
Our drivers run with ring sizes of 512 by default and configure the NIC to drop packets if the receive ring is full, a common setting to avoid buffer bloat.

\subsection{Test Setup}
Choice of language does not affect the average or median latency significantly:
garbage collection pauses and JIT compilation are visible in tail latency.
We therefore measure the latency of all forwarded packets by inserting fiber optic splitters on both sides of the device under test.
All packets are mirrored to a server running MoonSniff~\cite{moonsniff} with hardware timestamping on a Xeon D embedded NIC (measured precision 24\,ns, captures all packets).
The device under test uses an Intel Xeon E5-2620 v3 at 2.40\,GHz and a dual-port Intel X520 NIC.
All latencies were measured with a batch size of 32 and ring size 512 under bidirectional load with constant bit-rate traffic.
The device under test has a maximum buffer capacity of 1,088 packets in this configuration.
Different batch sizes, ring sizes, and NUMA configurations affect latency in the same way for all programming languages, interested readers are referred to the original publication about the C driver~\cite{ixy}.

\subsection{Tail Latencies}
\begin{figure}[t!]
\centering
    \subfigure[Forwarding latency at 1\,Mpps]{
        \hspace{-0.9em}\includegraphics[width=0.5\textwidth]{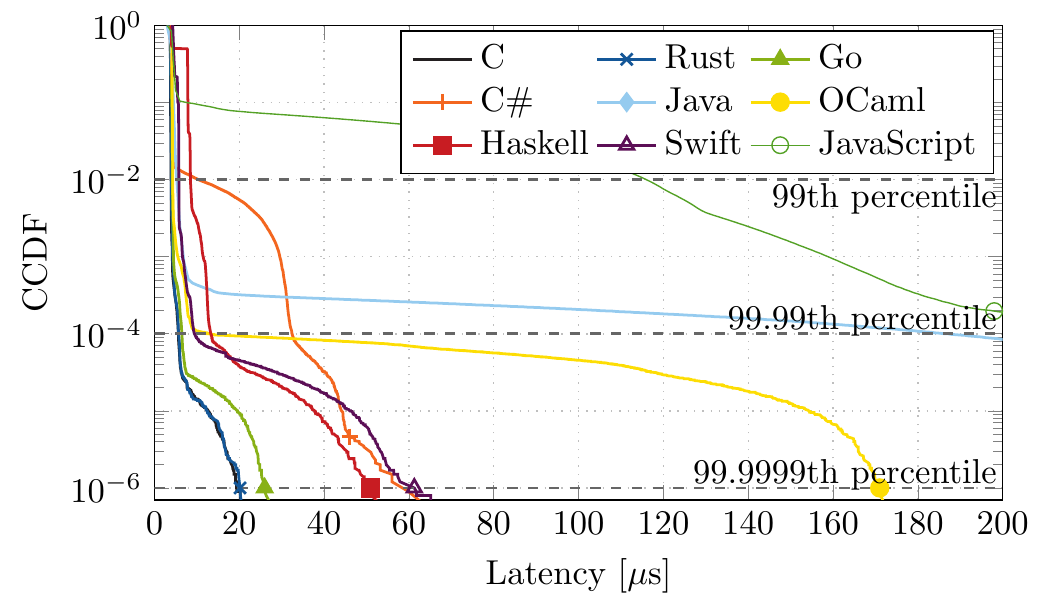}
        \label{fig:latency1}
    }
    \subfigure[Forwarding latency at 10\,Mpps]{
        \hspace{-0.9em}\includegraphics[width=0.5\textwidth]{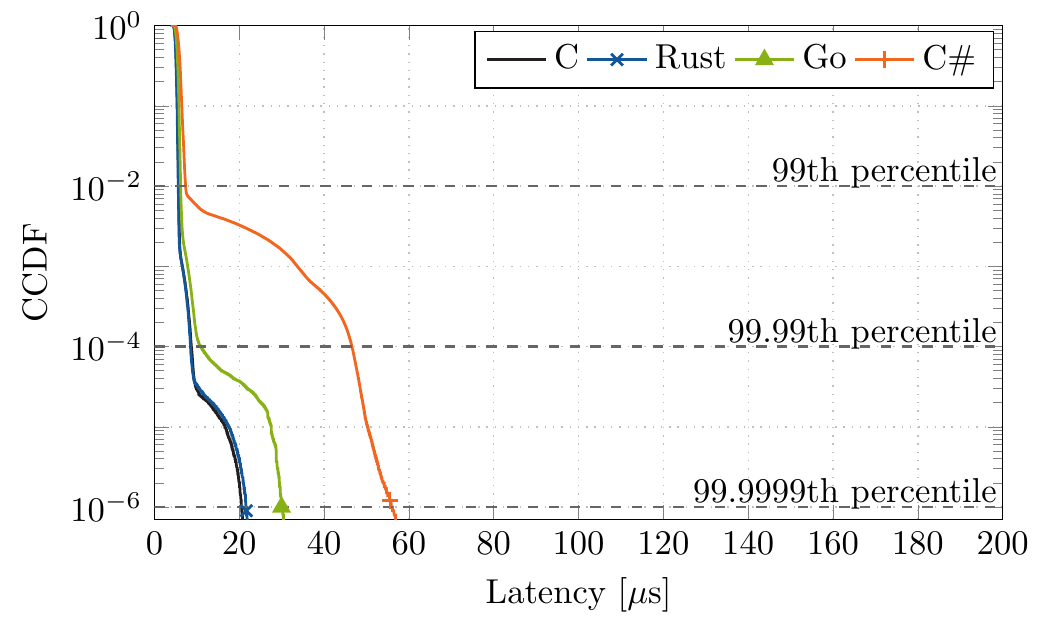}
        \label{fig:latency10}
    }
    \subfigure[Forwarding latency at 20\,Mpps]{
        \hspace{-0.9em}\includegraphics[width=0.5\textwidth]{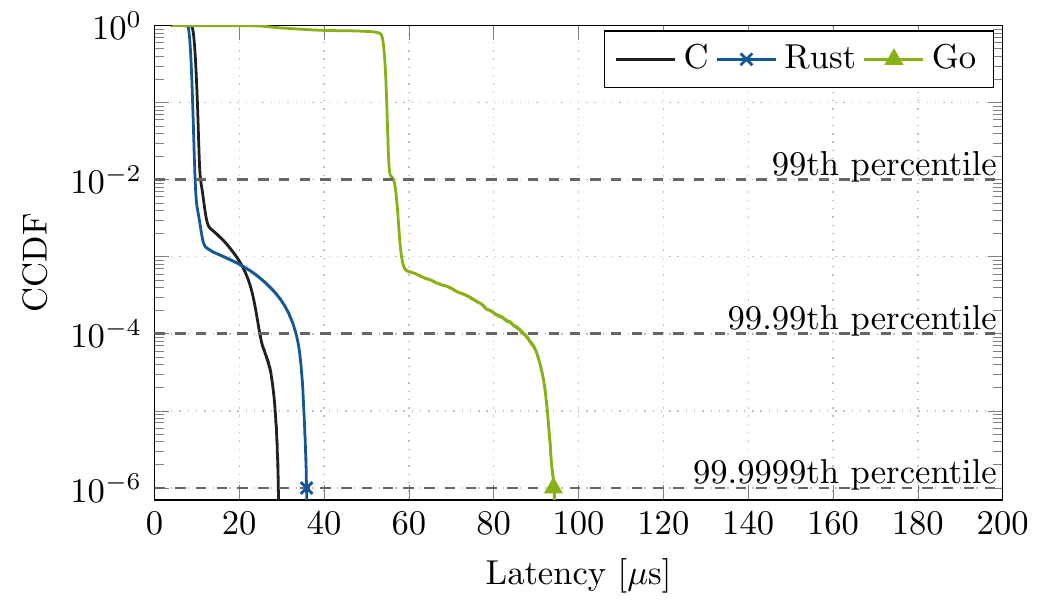}
        \label{fig:latency20}
    } 
    \caption{Tail latency of our implementations when forwarding packets}
    \label{fig:latency}
    \vspace{-1.5em}
\end{figure}

Figure~\ref{fig:latency} shows latencies of our drivers when forwarding packets at different rates.
The data is plotted as CCDF to focus on the worst-case latencies.
Implementations not able to cope with the offered load are omitted from the graphs --- their latency is simply a function of the buffer size as the receive buffers fill up completely.
No maximum observed latency is significantly different from the 99.9999th percentile. 
Java and JavaScript lose packets during startup due to JIT compilation, we therefore exclude the first 5 seconds of the test runs for these two languages.
All other tests shown ran without packet loss.

\subsubsection{Rust and C}
Even Rust and C show a skewed latency distribution with some packets taking 5 times as long as the median packet.
One reason for this is that all our drivers handle periodic (1\,Hz) printing of throughput statistics in the main thread.
Note that the 99.9999th percentile means that one in a million packets is affected.
Printing statistics once per second at 1\,Mpps or more thus affects latency at this level.
A second reason is that it is not possible to isolate a core completely from the system on Linux.
Some very short local timer interrupts are even present with the \texttt{isolcpus} kernel option.
C outperforms Rust at 20\,Mpps because Rust operates close to its limit of $\approx$24\,Mpps on this system and processes larger batches.

\subsubsection{Go}
Go's low-latency garbage collector achieves the lowest pause times of any garbage-collected language here.
Latency suffers at 20\,Mpps because the driver operates at its limit on this system here.
Cutler et al. measured a maximum garbage collection pause of 115\,$\mu$s in their Go operating system, demonstrating that sub-millisecond pauses with Go are possible even in larger applications.

\subsubsection{C\#}
C\# features several garbage collector modes tuned for different workloads~\cite{csharpgcs}.
The default garbage collector caused a latency of 240\,$\mu$s at the 99.9999th percentile at 10\,Mpps.
Switching it to \texttt{SustainedLowLatency} reduces latency to only 55\,$\mu$s, this change also reduces the maximum achievable throughput by 1.2\%.
All measurements were performed with the \texttt{SustainedLowLatency} garbage collector.

C\# also uses a JIT compiler that might introduce additional pause times.
However, the compilation of most functions happens immediately after startup even if no packets are forwarded: We implement a poll-mode driver that effectively warms up the JIT compiler.

\begin{figure}[t!]
	\vspace{0.5em}
	\hspace{-1.8em}\includegraphics[width=0.5\textwidth]{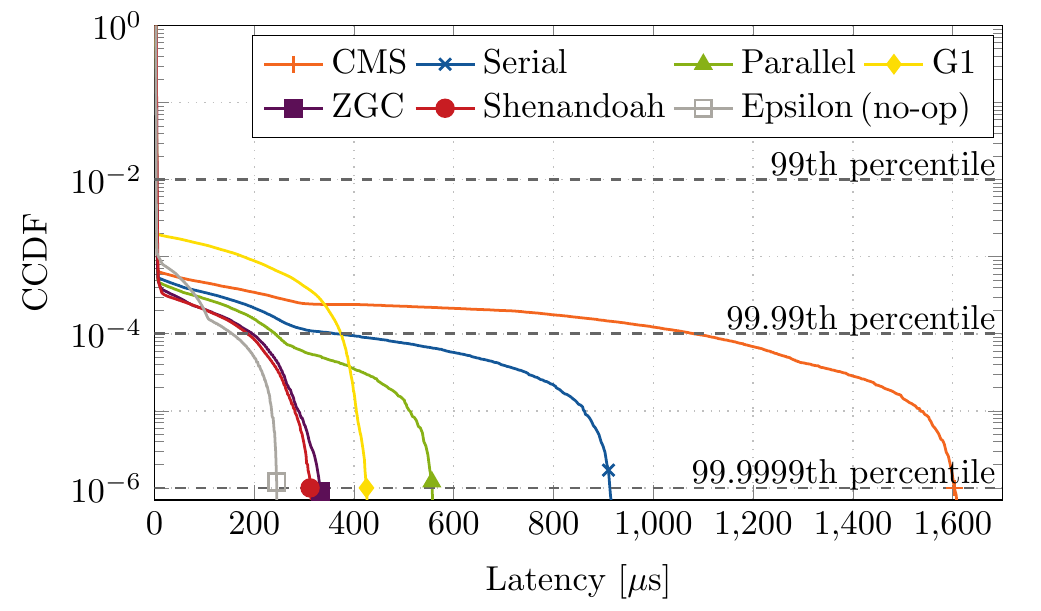}
	\caption{Forwarding latency of Java at 1\,Mpps with different garbage collectors}
	\label{fig:latency-java}
	\vspace{-1.5em}
\end{figure}

\subsubsection{Java}
Java exhibits packet loss and excessive latencies during the first few seconds of all test runs, this is likely due to JIT compilation hitting a function only after the traffic flow starts.
All latency measurements for Java therefore exclude the first 5 seconds of the test runs.
Figure~\ref{fig:latency1} shows results for the Shenandoah garbage collector which exhibited the lowest latency.
We also tried the different settings in Shenandoah that are recommended for low-latency requirements~\cite{shenandoah-tuning}.
Neither using a fixed-size heap with pre-touched pages, nor disabling biased locking made a measurable difference.
Changing the heuristic from the default \emph{adaptive} to \emph{static} reduces worst-case latency from 338\,$\mu$s to 323\,$\mu$s, setting it to \emph{compact} increases latency to 800\,$\mu$s.

Figure~\ref{fig:latency-java} compares the latency incurred by the different available garbage collectors in OpenJDK 12 while forwarding 1\,Mpps.
We configured the lowest possible target pause time of 1\,ms.
Note that the maximum buffer time with this configuration is $\approx$1.1\,ms, i.e., the CMS collector drops packets at this rate.
This could be mitigated by larger rings or by enabling buffering on the NIC if the ring is full.
There is a clear trade-off between throughput and latency for the different garbage collectors, cf. Table~\ref{tbl:java-gc-perf}.
ZGC hits a sweet spot between high throughput and low latency.
Even Epsilon (no-op, never frees objects) is also not ideal, indicating that the garbage collector is not the only cause of latency.
This can be attributed to the JIT and/or the bad data locality as it fills up the whole address space.

\subsubsection{OCaml and Haskell}
Both OCaml and Haskell ship with only a relatively simple garbage collector (compared to Go, C\#, and Java) not optimized for sub-millisecond pause times.
Haskell even drops packets due to garbage collection pauses when the multi-threaded runtime is enabled, the single threaded runtime performs reasonably well.
Haskell plans to provide a new low-latency garbage collector later in 2019~\cite{haskell-lowlatency-gc}.

\subsubsection{Swift}
It remains unclear why Swift performs worse than some garbage-collected languages. 
Its reference counting memory management should distribute the work evenly and not lead to spikes, but we observe tail latency comparable to the garbage-collected Haskell driver.

\subsubsection{JavaScript}
JavaScript loses packets during startup, indicating that the JIT compiler is to blame, the graph excludes the first 5 seconds.
The latency is still off the chart, the 99.99th percentile is 261\,$\mu$s, the 99.9999th percentile 353\,$\mu$s and the maximum 359\,$\mu$s.

\subsubsection{Python}
Python exhibits packet loss even at low rates and is therefore excluded here, worst-case latencies are several milliseconds even when running at 0.1\,Mpps.

\section{Conclusion}
Rewriting the entire operating system in a high-level language is a laudable effort but unlikely to disrupt the big mainstream desktop and server operating systems in the near future.
We propose to start rewriting drivers as user space drivers in high-level languages instead as they present the largest attack surface.
39 of the 40 memory safety bugs in Linux examined here are located in drivers, showing that most of the security improvements can be gained without replacing the whole operating system.
Network drivers are a good starting point for this effort: User space network drivers written in C are already commonplace.
Moreover, they are critical for security: they are exposed to the external world or serve as a barrier isolating untrusted virtual machines (e.g., CVE-2018-1059 in DPDK allowed VMs to extract host memory due to a missing bounds check~\cite{cve20181059}).

Higher layers of the network stack are also already moving towards high-level languages (e.g., the TCP stack in Fuchsia~\cite{fuchsia} is written in Go) and towards user space implementations.
The transport protocol QUIC is only available as user space libraries, e.g., in Chromium~\cite{chromium-quic} or CloudFlare's quiche written in Rust~\cite{quiche}.
Apple runs a user space TCP stack on mobile devices~\cite{appleuserspace}.
User space stacks also call for a more modern interface than POSIX sockets: the socket replacement TAPS is currently being standardized, it explicitly targets ``modern platforms and programming languages''~\cite{taps}.
This trend simplifies replacing the kernel C drivers with user space drivers in high-level languages as legacy interfaces are being deprecated.

Our evaluation shows that Rust is a prime candidate for safer drivers: its ownership-based management system prevents memory bugs even in custom memory areas not allocated by the language runtime.
The cost of these safety and security features are only 2\% - 10\% of throughput on modern out-of-order CPUs.
Rust's ownership based memory management provides more safety features than languages based on garbage collection here and it does so without affecting latency.
Linux kernel developers recently discussed the possibility to accept Rust code in the kernel as an optional dependency to enable safer code~\cite{rust-linux}.

Go and C\# are also a suitable language if the system can cope with sub-millisecond latency spikes due to garbage collection.
The other languages discussed here can also be useful if performance is less critical than having a safe and correct system, for example, Haskell and OCaml are more suitable for formal verification.

\section*{Reproducible Research}
We publish all of our source code and test scripts on GitHub at \href{https://github.com/ixy-languages/ixy-languages}{https://github.com/ixy-languages/ixy-languages}.
The only hardware requirement is a network card from the ixgbe family which are readily available from several vendors and often found as on-board NICs.

\section*{Acknowledgments}
This work was supported by the German-French Academy for the Industry of the Future.
We would like to thank Andreas Molzer, Boris-Chengbiao Zhou, Johannes Naab, Maik Luu Bach, Olivier Coanet, Sebastian Gallenmüller, Stefan Huber, Stephan-Alexander Posselt, and Thomas Zwickl for valuable contributions to the driver implementations and/or this paper.

\vfill
\pagebreak

\appto\UrlNoBreaks{\do\:}
\bibliographystyle{ACM-Reference-Format}
\renewcommand*{\bibfont}{\fontsize{10}{12}\selectfont}
{\large\bibliography{lit}}

\end{document}